\documentclass[useAMS,usenatbib]{mn2e}

\newcommand{\etal}{\mbox{et\ al.}}
\newcommand{\ang}{\,\mbox{\AA}}

\newcommand{\msun}{\,\mbox{$\mbox{M}_{\odot}$}}

\newcommand{\aanda}  {A\&A,~\nolinebreak }
\newcommand{\aasupp} {A\&AS,~\nolinebreak}

\newcommand{\mn}     {MNRAS,~\nolinebreak }
\usepackage{rotating}
\usepackage{amssymb}

\title[Spectroscopy of GRB\,021004]
{Spectroscopy of GRB\,021004: a structured jet ploughing through a massive stellar wind}

\author[Starling \etal]
{R. L. C. Starling$^1$, R. A. M. J. Wijers$^1$, M. A. Hughes$^2$, N. R. Tanvir$^2$, \and\
P. M. Vreeswijk$^3$, E. Rol$^4$ \& I. Salamanca$^1$
\\$^1$Astronomical Institute `Anton Pannekoek', University of Amsterdam, Kruislaan 403, 1098 SJ Amsterdam, The Netherlands\\$^2$Centre for Astrophysics Research, STRI, University of Hertfordshire, College Lane, Hatfield, Herts. AL10 9AB, UK\\$^3$European Southern Observatory, Alonso de C\'{o}rdova 3107, Casilla 19001, Santiago 19, Chile\\$^4$Dept. of Physics and Astronomy, University of Leicester, University Road, Leicester LE1 7RH, UK}

\begin{document}
\date{Accepted . Received ; in original form 2005 January 06}

\pagerange{\pageref{firstpage}--\pageref{lastpage}} \pubyear{2005}

\maketitle

\label{firstpage}


\begin{abstract}
We present spectra of the afterglow of GRB\,021004 taken with WHT ISIS and VLT FORS1 at three epochs spanning 0.49--6.62 days after the burst. We observe strong absorption likely coming from the host galaxy, alongside absorption in {\small H\,I}, {\small Si\,IV} and {\small C\,IV} with blueshifts of up to 2900~km~s$^{-1}$ from the explosion centre which we assume originates close to the progenitor. We find no significant variability of these spectral features. We investigate the origin of the outflowing material and evaluate various possible progenitor models. The most plausible 
explanation is that these result in the fossil stellar wind of a highly
evolved Wolf-Rayet star. However, ionization from the burst itself prevents the existence of {\small H\,I}, {\small Si\,IV} and {\small C\,IV} close to the afterglow surface where the fast stellar wind should dominate, and large amounts of blueshifted hydrogen are not expected in a Wolf-Rayet star wind. We propose that the Wolf-Rayet star wind is enriched by a hydrogen-rich companion, and that the GRB has a structured jet geometry in which the gamma rays emerge in a small opening angle within the wider opening angle of the cone of the afterglow. This scenario is able to explain both the spectral line features and the irregular light curve of this afterglow.   
\end{abstract}


\section{Introduction}
The location of Gamma-ray bursts (GRBs) at cosmological distances was settled by the measurement of absorption lines in optical afterglow spectra, first done for GRB\,970508 \citep{Metzger}. Many bursts are now followed up spectroscopically, not only for redshift determination but also as a probe of conditions in the host galaxy, the circumburst environment and intervening material. Spectral features imprinted on the afterglow spectrum can be a useful tool for understanding Gamma-ray Burst origins. In several, and at least 2, cases afterglow spectroscopy, sometimes in combination with photometry, has revealed the presence of a supernova connected with the GRB (e.g. GRB\,980425 and SN1998bw, Galama \etal\ 1998; GRB\,011121 and SN2001ke, Garnavich \etal\ 2003; GRB\,030329 and SN2003dh, Hjorth \etal\ 2003; Stanek \etal\ 2003 and GRB\,031203 and SN2003lw, Malesani \etal\ 2004).  
 
\begin{figure}
\begin{center}
\includegraphics[width=6cm, angle=90]{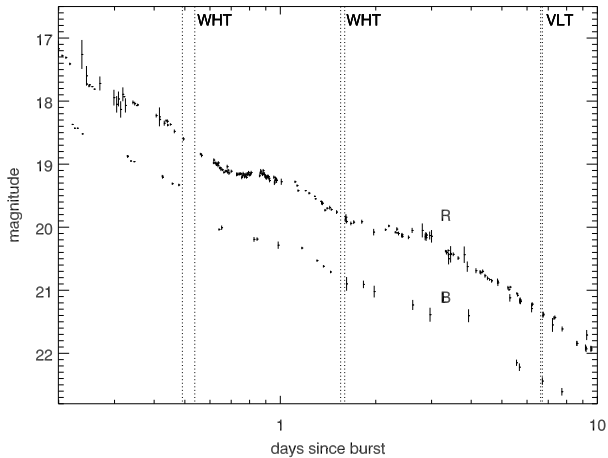}
\caption{$B$ and $R$ band photometry of GRB\,021004 compiled from the literature (Bersier \etal\ 2003, Fox \etal\ 2003, Holland \etal\ 2003, Pandey \etal\ 2003 and Uemura \etal\ 2003). The coverage of the WHT and VLT spectra presented here is shown by vertical dotted lines.}
\label{photometry}
\end{center}
\end{figure}

GRB\,021004 was detected on 2002 October 4, 12:06 UT by the HETE-2 satellite \citep{discovery}. An optical afterglow was discovered ten minutes after detection of the prompt gamma-ray emission with an $R$ band magnitude of 15.34 \citep{Foxa}. Intervening absorption systems at $z=1.38$ and $z=1.60$ were found in the subsequent afterglow spectra \citep{Foxb}, and later spectra showed further absorption features at redshift $z=2.323$ taken to be the host galaxy redshift \citep{redshift}. The afterglow light curve is not simply a power law in shape, but shows several bumps, particularly in the well sampled region around 1 day after the burst (Fig. \ref{photometry}). The temporal decay at early times (before $\sim$ 1 day) follows what is expected for a jet expanding into a stellar wind medium with a 1/r$^{2}$ dependence \citep{LiChev}, but when including the later times several different explanations are possible. 

GRB\,021004 was localised in the optical band very early, allowing unusually prompt spectroscopic follow-up. The optical afterglow is bright, and rich in line features including {\small Si\,IV} and {\small C\,IV}. It is the first afterglow for which the spectrum is dominated by absorption lines spanning a 3000~km~s$^{-1}$ velocity range, first reported by Salamanca \etal\ (2002). A number of high redshift afterglow spectra show absorption features from {\small Si\,IV} and {\small C\,IV} in their spectra, including GRB\,000926 \citep{000926}, GRB\,011211 \citep{Holland02} and GRB\,030323 (including an outflowing system with $v = 130\pm60$ km s$^{-1}$, Vreeswijk \etal\ 2004). Such lines are also common in the interstellar medium (ISM) of many galaxies and also in a composite spectrum of $z\sim3$ Lyman Break galaxies \citep{LBGs}, but in these cases never reach the high velocities observed in the afterglow of GRB\,021004. Despite many detailed studies of afterglow spectra of GRB\,021004, the origin of the spectral features remains unclear.

We present early afterglow spectra taken with the 4.2-m William Herschel Telescope (WHT) and late-time spectra taken with the 8.2-m Very Large Telescope (VLT). Spectra were obtained 11.78 hours after the burst - the earliest optical spectra of GRB\,021004 taken by a 4-m class telescope, and also at 1.55 and 6.62 days after the burst.
We analyse in detail the multiple absorption complexes at all three epochs, and investigate their origins. 
\begin{figure*}
\begin{center}
\includegraphics[width=11cm, angle=90]{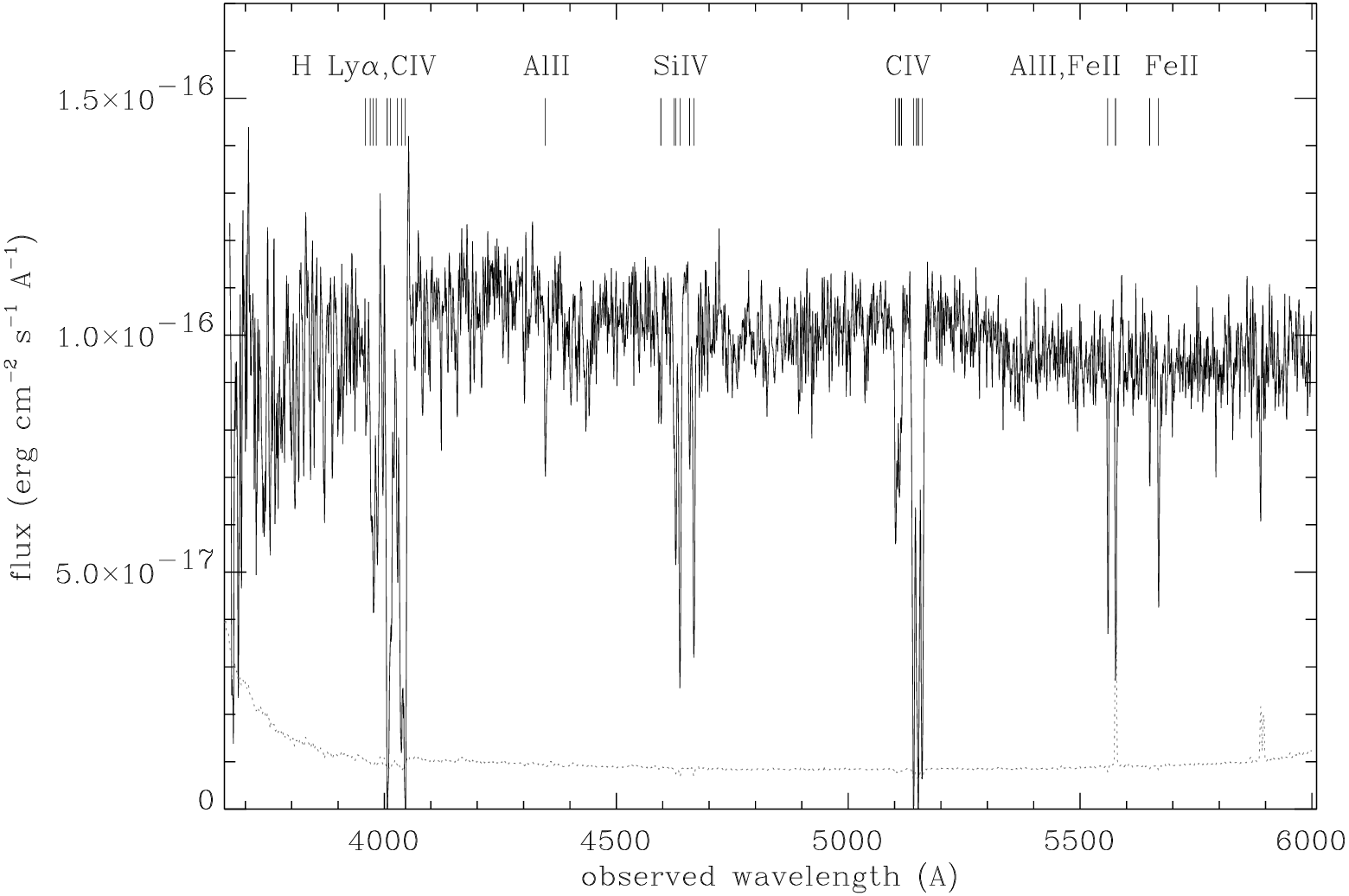}
\caption{The WHT blue grating spectrum at epoch 1 (solid line) and the error spectrum (dashed line), fluxes corrected according to photometric light curve in Fig. 1. Identified lines are labelled above.}
\label{spectrum1}
\end{center}
\end{figure*}
  
\section{Observations}
The WHT spectra were taken on 2002 October 4, 23:52:60 UT and October 6, 01:23:55 UT with the ISIS spectrograph. Overlapping blue (B300B) and red (R316R+GG495 filter) gratings were used with the EEV12 and Marconi2 CCDs respectively. Two exposures totalling 4000 s at epoch 1 and 4800 s at epoch 2 were taken with each grating, and with slits of width 0.87 arcsec at epoch 1 and 1.03 arcsec at epoch 2.

Flat fields were taken during the first night, and CuNe+CuAr arc lamps were obtained before and after each set of target exposures. The standard stars SP0305+261 and SP2323+157 were observed directly preceding or following each set of target and arc exposures on both nights. The airmass ranged between 1.01 and 1.09 and conditions were not photometric. We applied a correction for atmospheric extinction using the mean extinction curve for La Palma, and a Galactic extinction correction using $E(B-V)$ = 0.060 as determined by Schlegel, Finkbeiner \& Davis (1998). 
Data reduction was performed using {\small IRAF}, except for cosmic ray removal, where we used the {\small L. A. Cosmic} routine to remove both point source and irregular-shaped cosmic ray traces \citep{lacosmic}. For each night, we combine the individual exposures of GRB\,021004 to achieve a higher signal to noise.

The useful remaining wavelength ranges are 3650--6000\ang\ for the blue grating and 6500--8500\ang\ for the red grating. The wavelength resolution is approximately 2.2\ang\ at epoch 1 and 3.5\ang\ at epoch 2. 
The average signal to noise is 12.7 at epoch 1, 7.6 at epoch 2.

A third epoch observation was made with the Very Large Telescope (VLT-Antu) in Paranal on 2002 October 11, 02:57:09 UT. The observations were made with the Focal Reducer/Low Dispersion Spectrograph 1 (FORS1) instrument in the Longslit Spectroscopy (LSS) mode.
Six 1200 s exposures were taken through a 0.7 $\times$ 416.8~arcsec slit, with the 600B grism (3450--5900\AA).
Data reduction was carried out using a combination of routines within the {\small IRAF} and Starlink {\small FIGARO} packages. The 6 object images were combined into a single spectrum. Wavelength calibration was perfomed using He and HgCd lamps, and the standard star LTT~7987 was used for flux calibration with airmasses ranging from 1.38--1.57. The resulting spectrum has a signal to noise ratio of 4.6 at the longest wavelengths down to $\sim$~2.5 at H Ly$\alpha$, and a wavelength resolution of $\sim$ 3.8\ang.

Spectral analysis was done using the Starlink {\small DIPSO} spectral fitting package. To obtain the correct continuum fluxes, the data are compared with $B$ and $R$ band photometry collected from the literature (Fig. \ref{photometry}). The photometric points closest to the observation dates were interpolated to give the magnitudes $B_{\rm epoch1}\sim19.50$, $B_{\rm epoch2}\sim20.83$ and $R_{\rm epoch1}\sim18.55$
expected in the WHT spectra and $B\sim22.44$ at the time of the spectroscopic VLT observation. 
The observed fluxes in the WHT spectra are as expected at epoch 2, but the earlier spectra are fainter by a factor 2.2 for the blue grating and 1.7 for the red grating. The flux in the VLT spectrum is 1.6 times lower than expected from photometric measurements.
\subsection{Spectral line fitting}
We show the first epoch WHT ISIS spectrum and its error spectrum in Fig. \ref{spectrum1} with all identified lines labelled. We fit a local continuum around each line complex with a first order polynomial. Each doublet was modelled with two Gaussian profiles, with the line full widths at half maxima (FWHM)
of the two components tied. For each species, all doublet fits (fit parameters of FWHM, central wavelength and peak intensity) were optimised simultaneously using the least squares method. The H Ly$\alpha$ region has to be treated differently as individual lines are much less clear, so here we fitted a number of broad absorption troughs representing some unknown number of blended lines, plus the narrow emission line.
The error on line equivalent widths was measured within the {\small DIPSO} package using the statistical uncertainties in the continuum fit and on the flux values within the line, and also the systematic error in continuum placement and zero level in the case of the WHT data \citep{dipso}.

\begin{table*}
\tiny
\caption{Model fits to the identified spectral line complexes at both epochs. Velocities and equivalent widths are in the restframe of the source. EW errors are quoted at the 2$\sigma$ level and EWs marked with a $^{*}$ indicate lower limits for saturated lines. $^{+}$ indicates the two heavily blended lines of {\small C\,IV} for which only the total EW should be considered. Velocities and redshifts are given to 1$\sigma$ accuracy and precision of observed wavelengths is limited by the wavelength resolutions specified in Section 2. Epochs 1 and 2 refer to WHT ISIS data, epoch 3 refers to VLT FORS1 data.}
\begin{tabular}{lllllllll}
\hline
$\lambda_{\rm obs\rm }$ (\AA) & Line & epoch&$z$&$v_{\rm internal}$ FWHM (km~s$^{\rm -1}$)& EW (\AA)&$v_{\rm outflow}$ (km~s$^{-1}$)&system\\ \hline \hline
3959,3969,3976,3982&see text for ID&1&&$<$350$>$ & 2.44$\pm$0.32&&\\
3959,3969,3976,3982&&2&& $<$330$>$& 2.11$\pm$0.70&&\\ \hline \hline
4006&Ly$\alpha$ $\lambda$1215 \AA&1& 2.297$\pm$0.002&410$\pm$45&1.52$\pm$0.27$^{*}$& 2700$\pm$170&IV\\
4005&&2& 2.296$\pm$0.003&450$\pm$250&1.28$\pm$0.62$^{*}$& 2800$\pm$270&\\
4008&&3& 2.297$\pm$0.003&360$\pm$40&2.37$\pm$0.87$^{*}$& 2700$\pm$290&\\ \hline
4013&Ly$\alpha$ $\lambda$1215 \AA&1& 2.303$\pm$0.002&670$\pm$140&1.96$\pm$0.35& 2200$\pm$170&III\\
4011&&2& 2.301$\pm$0.003&620$\pm$450&2.25$\pm$1.02& 2400$\pm$270&\\ \hline \hline
4028&C IV $\lambda$1548 \AA&1& 1.602$\pm$0.001& 240$\pm$70& 0.60$\pm$0.16& intervening &\\
4026&&2& 1.601$\pm$0.002&230$\pm$60&0.71$\pm$0.42&intervening&\\
4028&&3& 1.602$\pm$0.003&240$\pm$40&1.54$\pm$0.99&intervening &\\ \hline \hline
4037&Ly$\alpha$ $\lambda$1215 \AA&1& 2.323$\pm$0.001&740$\pm$160&2.83$\pm$0.48& 360$\pm$170&II\\
4036&&2& 2.322$\pm$0.003&690$\pm$100&2.48$\pm$1.12& 450$\pm$270&\\
4037&&3& 2.321$\pm$0.003&400 fixed&2.09$\pm$1.01&560$\pm$280 &\\ \hline
4045&Ly$\alpha$ $\lambda$1215 \AA&1& 2.329$\pm$0.002&350$\pm$80&1.39$\pm$0.25$^{*}$& -180$\pm$170& I\\
4044&&2& 2.328$\pm$0.003&280$\pm$70&0.97$\pm$0.49$^{*}$& -90$\pm$270&\\ 
4043&&3& 2.326$\pm$0.003&220$\pm$40 &0.65$\pm$0.70$^{*}$& 50$\pm$280&\\ \hline \hline
4346&Al II $\lambda$1670 \AA&1& 1.602$\pm$0.001& 350$\pm$80&0.68$\pm$0.15&intervening&\\
4346&&2& 1.602$\pm$0.002& 280$\pm$70&0.57$\pm$0.31&intervening&\\ \hline \hline
4596&Si IV $\lambda$1393 \AA&1& 2.297$\pm$0.002 &230$\pm$70&     0.23$\pm$0.09&2700$\pm$150&IV\\
4596&&2& 2.297$\pm$0.003 &110$\pm$40&     0.13$\pm$0.13&2700$\pm$230&\\
4598&&3&2.298$\pm$0.003& 60$\pm$20& 0.28$\pm$      0.39&2600$\pm$250& \\
4622&Si IV $\lambda$1402 \AA&1&      2.296$\pm$0.002 &230$\pm$70&     0.26$\pm$0.10&2800$\pm$150&\\
4624&&2&      2.295$\pm$0.003  &110$\pm$40&     0.14$\pm$0.12&2900$\pm$230&\\ 
4623&&3&      2.296$\pm$0.003&150$\pm$30& 0.51$\pm$0.42&2800$\pm$250& \\\hline
4628&Si IV $\lambda$1393 \AA&1&      2.320$\pm$0.002&220$\pm$20&     0.57$\pm$0.12&630$\pm$150&II\\
4628&&2&      2.320$\pm$0.003  &350$\pm$50&     0.96$\pm$0.40&630$\pm$230&\\
4629&&3&2.321$\pm$0.003&290$\pm$60&0.56$\pm$0.48&520$\pm$250 &\\
4658&Si IV $\lambda$1402 \AA&1&      2.320$\pm$0.002 &220$\pm$20&     0.33$\pm$0.10&630$\pm$150&\\
4658&&2&      2.320$\pm$0.003  &340$\pm$50&     0.41$\pm$0.22&630$\pm$230&\\ 
4660&&3&2.322$\pm$0.003&290$\pm$60&0.26$\pm$0.52&430$\pm$240 &\\\hline
4637&Si IV $\lambda$1393 \AA&1&      2.327$\pm$0.002 &300$\pm$10&      1.13$\pm$0.20&  0$\pm$150&I\\
4637&&2&      2.326$\pm$0.003  &270$\pm$20&      1.05$\pm$0.42&90$\pm$230&\\
4637&&3&2.327$\pm$0.003&250$\pm$20&1.38$\pm$0.51&-20$\pm$250 &\\
4667&Si IV $\lambda$1402 \AA&1&      2.327$\pm$0.002 &300$\pm$10&      1.04$\pm$0.19&  0$\pm$150&\\
4667&&2&      2.326$\pm$0.003  &260$\pm$20&      1.03$\pm$0.41&90$\pm$230&\\ 
4668&&3& 2.327$\pm$0.003& 250$\pm$20&1.11$\pm$0.51& -40$\pm$240 &\\ \hline \hline
5102&C IV $\lambda$1548 \AA&1&      2.296$\pm$0.001 &230$\pm$20&     0.60$\pm$0.11  & 2800$\pm$130&IV\\
5102&&2&      2.296$\pm$0.002 &310$\pm$60&     0.60$\pm$0.26&2800$\pm$210&\\
5102&&3&2.295$\pm$0.003& 270$\pm$30&0.71$\pm$0.12& 2900$\pm$230 &\\
5106&C IV $\lambda$1550 \AA&1&      2.296$\pm$0.001 &230$\pm$20&     0.39$\pm$0.08&2800$\pm$130&\\
5106&&2&      2.296$\pm$0.002 &310$\pm$60&     0.51$\pm$0.23&2800$\pm$210&\\ 
5108&&3&2.294$\pm$0.003& 270$\pm$30&0.30$\pm$0.11& 3000$\pm$230&\\ \hline
5111&C IV $\lambda$1548 \AA&1&      2.302$\pm$0.001 &190$\pm$20&     0.38$\pm$0.08&2300$\pm$130&III\\
5111&&2&      2.302$\pm$0.002 &120$\pm$20&     0.20$\pm$0.12&2300$\pm$210&\\
5113&&3&2.302$\pm$0.003&100$\pm$20 &0.27$\pm$0.09& 2300$\pm$220&\\
5114&C IV $\lambda$1550 \AA&1&      2.300$\pm$0.001 &190$\pm$20&     0.23$\pm$0.07&2500$\pm$130&\\
5114&&2&      2.300$\pm$0.002 &120$\pm$20&     0.21$\pm$0.12&2500$\pm$210&\\ 
5117&&3&2.300$\pm$0.003&100$\pm$20 &0.23$\pm$0.09&2500$\pm$220 &\\ \hline
5141&C IV $\lambda$1548 \AA&1&      2.321$\pm$0.001 &340$\pm$10&      1.79$\pm$0.28$^{*}$&540$\pm$130&II\\
5140&&2&      2.320$\pm$0.002 &330$\pm$100&      1.53$\pm$0.57$^{*}$&630$\pm$210&\\
5141&&3&2.321$\pm$0.003& 220$\pm$20&1.68$\pm$0.13$^{*}$& 670$\pm$220&\\
5149&C IV $\lambda$1550 \AA$^{+}$&1&      2.321$\pm$0.001 &330$\pm$10&1.06$\pm$0.17$^{*}$&540$\pm$130&\\
5148&&2&      2.319$\pm$0.002 &330$\pm$100&0.64$\pm$0.27$^{*}$&720$\pm$210&\\ 
5150&&3&2.321$\pm$0.003&300$\pm$10 &1.90$\pm$0.12$^{*}$&900$\pm$220 &\\ \hline
5152&C IV $\lambda$1548 \AA$^{+}$&1&      2.328$\pm$0.001 &310$\pm$10&1.28$\pm$0.21$^{*}$&  -90$\pm$130&I\\
5150&&2&      2.327$\pm$0.002 &340$\pm$30&2.09$\pm$0.77$^{*}$& 0$\pm$210&\\
5156&&3&2.330$\pm$0.003&300$\pm$10 &0.48$\pm$0.11$^{*}$& 50$\pm$220&\\
5160&C IV $\lambda$1550 \AA&1&      2.329$\pm$0.001 &310$\pm$10&      1.64$\pm$0.26$^{*}$& -180$\pm$130&\\
5159&&2&      2.329$\pm$0.002 &340$\pm$30&      1.89$\pm$0.70$^{*}$&  -180$\pm$210&\\ 
5160&&3&2.328$\pm$0.003& 300$\pm$10&1.31$\pm$0.11$^{*}$& -20$\pm$220&\\ \hline \hline

5559&Al II $\lambda$1670 \AA&1&2.327$\pm$0.001&160$\pm$10&0.59$\pm$0.13& 0$\pm$120&I\\
5559&&2&2.327 fixed&100$\pm$30&0.19$\pm$0.16& 0$\pm$190&\\ \hline \hline
5576,5650,5669&Fe II $\lambda$2344,2374,2382 \AA&1&1.380$\pm$0.001&$<$160$\pm$20$>$&2.25$\pm$0.33&intervening &\\
5576,5650,5669&&2&1.380$\pm$0.001&$<$100$\pm$30$>$&2.34$\pm$0.73&intervening&\\ \hline \hline
6654&Mg II $\lambda$2796 \AA&1&1.380$\pm$0.001& 230$\pm$10&1.52$\pm$0.26& intervening &\\
6671&Mg II $\lambda$2803 \AA&1&1.380$\pm$0.001& 250$\pm$20&1.23$\pm$0.22& intervening &\\ \hline \hline
6731&Fe II $\lambda$2586 \AA&1& 1.603$\pm$0.001&130$\pm$30&  0.34$\pm$0.10& intervening& \\
6765&Fe II $\lambda$2599 \AA&1& 1.603$\pm$0.001&220$\pm$20&  0.74$\pm$0.16& intervening& \\ \hline \hline
7278&Mg II $\lambda$2796 \AA&1& 1.603$\pm$0.001&220$\pm$70& 1.27$\pm$0.23& intervening &\\
7296&Mg II $\lambda$2803 \AA&1& 1.603$\pm$0.001&210$\pm$40& 1.54$\pm$0.27& intervening &\\ \hline \hline
\end{tabular}
\label{tab:lineparams}
\end{table*}
\normalsize

\begin{figure}
\begin{center}
\includegraphics[width=8cm]{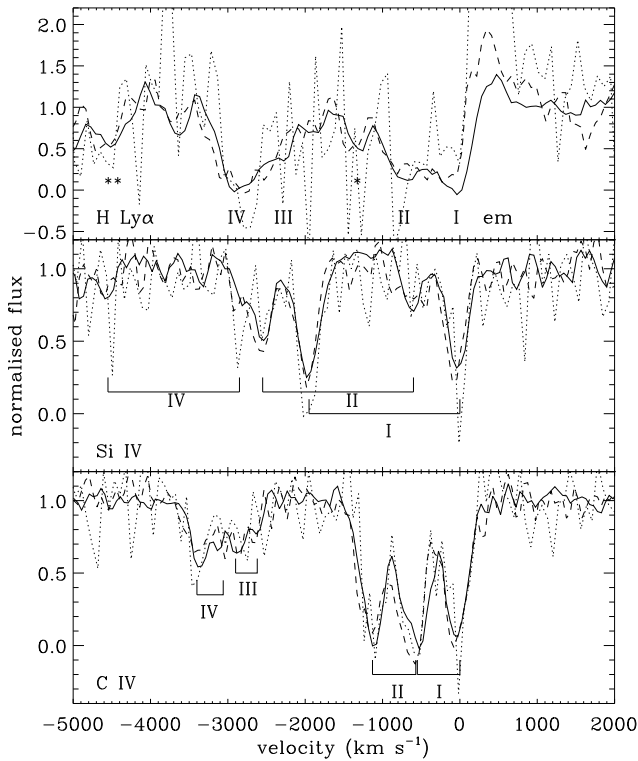}
\caption{Outflow velocities of the {\small C\,IV} and {\small Si\,IV} absorption doublets (velocities plotted are for the longer wavelength components only) and corresponding hydrogen Ly$\alpha$ systems, divided by the local continua. The solid lines show the WHT observation of 4/10/02, the dashed lines show the WHT observation of 6/10/02 and the dotted lines show the VLT observation of 11/10/02. The 4 velocity systems are denoted by roman numerals, the Ly$\alpha$ emission line (too large in the VLT data to be shown fully on this plot) is labelled `em'. Intervening {\small C\,IV} is marked with a * and ** marks the long-wavelength end of the bluemost absorption complex.}
\label{velocity}
\end{center}
\end{figure}

\section{Results} 
The results of spectral fitting are detailed in Table \ref{tab:lineparams}.
We observe H Ly$\alpha$, {\small Si\,IV} and {\small C\,IV} absorption features at 3 redshifts, and H Ly$\alpha$ and {\small C\,IV} at a further redshift (this is the weakest velocity system and is not detected in {\small Si\,IV} or Ly$\alpha$ at epoch 3), all around $z\sim$~2.3. Other than the intervening systems, all absorption lines have a component at $z=2.327$, the highest redshift absorption system we observe, which we take to be the redshift of the host galaxy.
If the other 3 systems are a type of outflow (we do not rule out the possibility that the 0 velocity system is associated with the GRB progenitor rather than the host galaxy, but see Section 4.2), their observed velocities would be approximately 570, 2400 and 2800~km~s$^{-1}$ (as reported in Salamanca \etal\ 2002). 

The lines at we observe at 1190--1197\AA\ could be either
H Ly$\alpha$ at 5500 km~s$^{-1}$, intervening Ly$\alpha$ forest lines or {\small Si\,II} at 0 km~s$^{-1}$ ($z=2.327\pm0.003$). In the former case, we might expect
to see {\small C\,IV} or {\small Si\,IV} at this velocity as well, which we do not; in the latter case,
the stronger {\small Si\,II $\lambda$1260} line should also be seen, but it is not.
We conclude that the nature of the lines at 1190--1197\AA\ is uncertain,
though most likely Ly$\alpha$ forest lines, so we claim no H at velocities above 3000 km~s$^{-1}$.

\subsection{Variability}
We find no significant variability in equivalent width (EW, within the 2$\sigma$ errors) between the absorption features common to 2 or more epochs, with the exception of {\small Al\,II $\lambda$1670} at $z$ = 2.327 which we were unable to clearly detect in the 2nd and 3rd epochs. Fixing the position of {\small Al\,II} to that found at epoch 1 allowed us to fit a line there, and the resulting EW is inconsistent with that at epoch 1 within the 2$\sigma$ errors. The {\small Al\,II} line is also detected in several other spectra of this afterglow (M\o ller \etal\ 2002; Matheson \etal\ 2003; Mirabal \etal\ 2003; Schaefer \etal\ 2003; Fiore \etal\ 2004) taken at times both before (in the case of the 2.56-m Nordic Optical Telescope (NOT) data) and after our 2nd epoch WHT spectra, and in each case the measured EW is consistent with what we measure in the epoch 1 WHT data. The discrepancy may be due to imperfect cosmic ray/bad column removal at epoch 2, and/or there may be moderate variability in this line. {\small Fe\,II $\lambda$2344} is another low ionization line which is not detected in all optical spectra of GRB\,021004, though we do find it in the first 2 epochs presented here. The apparent variability of this line is largely due to the presence of a strong skyline near that position. The lack of significant variability in the strong lines indicates that the absorbing material in our line-of-sight remains in front of the afterglow throughout the $\sim$ 6 days spanned by our spectra.

The EWs we measure are consistent with those of corresponding lines or line complexes reported by M\o ller \etal\ (2002), Matheson \etal\ (2003), Mirabal \etal\ (2003) and Fiore \etal\ (2004).
The resolution of these spectra are such that we are not resolving every component of each line blend -- the line complex at the host galaxy redshift is resolved into 3 components in the VLT UVES spectrum \citep{fiore}. Unfortunately the calibration uncertainties in our late-time spectrum are too large to detect the colour change blueward of Ly$\alpha$ of $\sim0.3$ magnitudes claimed by Matheson \etal\ (2003) and Schaefer \etal\ (2003) and supported by photometric observations \citep{Bersier}, but not found in the HST observations of Fynbo \etal\ (2004).

\subsection{Hydrogen Ly$\alpha$ emission}
We clearly detect the host galaxy ISM through its emission in hydrogen Ly$\alpha$. Taking the host galaxy redshift to be $z=2.327$,
the Ly$\alpha$ emission line centre then appears redshifted, by an amount that is somewhat different between
various authors. However, as it lies right on the edge of the Ly$\alpha$
absorption complex, which probably eats into the emission line, its
central wavelength will depend on the resolution of the spectrum, so
this effect is not surprising. Within the resolution of these spectra (a few hundred km s$^{-1}$),
its velocity is in fact consistent with zero.

The Ly$\alpha$ emission line strength is difficult to estimate due to the line blending. However, by fitting in isolation the part of the line observed above the continuum level, we can measure a lower limit to the flux:\\
F$_{\rm epoch~1} \ge 2.35\pm0.87\times10^{-16}$ erg cm$^{-2}$ s$^{-1}$, F$_{\rm epoch~2} \ge 1.66\pm0.34\times10^{-16}$ erg cm$^{-2}$ s$^{-1}$. These lower limits are consistent within the quoted 1$\sigma$ errors, suggesting that both the emission line and the blended absorption line component at the host redshift show no variability, and are consistent with the value measured in NOT data of this source \citep{moller}. In epoch 3 we obtain F$_{\rm epoch~3} \ge 1.17\times10^{-16}$ erg cm$^{-2}$ s$^{-1}$, with a large error of $\sim$ 56 per cent.

Following the method described in Fynbo \etal\ (2004, and references therein) we calculate a star formation rate (SFR) using the Ly$\alpha$ emission line flux measured in the WHT data. 
We find a lower limit for the SFR of 6 \msun~yr$^{-1}$, with average values of 10 and 7 \msun~yr$^{-1}$ at epochs 1 and 2 respectively, consistent with estimates based on Keck and HST data of this afterglow (Djorgovski \etal\ 2002; Fynbo \etal\ 2004). Our estimate assumes a cosmology with H$_0$ = 70~km~s$^{-1}$~Mpc$^{-1}$, $\Omega_{\rm M}$ = 0.3 and $\Omega_{\Lambda}$ = 0.7.

From these data we can obtain only a lower limit for the column density of {\small H\,I} of $N_{\rm HI}\ge 10^{14}$ cm$^{-2}$, by comparing the observed EWs of each line to their theoretical curves of growth. This poor constraint is because the {\small H\,I} lines are close to or at saturation, they are all heavily blended and cannot be well modelled. The column of neutral hydrogen is in fact much larger: $1\times10^{18} \le N_{\rm HI}$ (cm$^{-2}$) $\le 1.1\times10^{20}$ (M\o ller \etal\ 2002; Fynbo \etal\ 2004). From the unsaturated lines of {\small Si\,IV} and {\small C\,IV} (those at the 2 lowest redshifts and part of the outflow) we measure columns of $\log N = 14.0\pm0.9$ and $14.6\pm0.9$ respectively, where $N$ is given in cm$^{-2}$, in line with previous estimates (Schaefer \etal\ 2003 {\small C\,IV} only; Fiore \etal\ 2004).

\section{Discussion}
The spectra we have taken provide information on material whose distance
from the explosion centre covers a wide range. No material can be closer
than the emitting surface (the blast wave) at the time of the first
spectrum, which is about $10^{16}$~cm.
Some lines are from intervening 
galaxies very far from the host at $z=1.38$ and $1.60$, and yet others
are from the host itself at $z=2.327$. We interpret the low-ionization
absorption lines and the narrow Ly$\alpha$ emission line
around $z\simeq2.33$ as coming from within the host, but many parsecs from
the explosion so that the absorbing material was unaffected by the
explosion or by the GRB progenitor. The blueshifted
lines we interpret as coming from the region affected by the explosion or
the progenitor, which we shall term the `high-velocity zone'. 
The Ly$\alpha$ absorption could in principle come from
either the high-velocity zone or the host, but given its velocity structure, identical to that of the high-ionization lines of {\small Si\,IV} and {\small C\,IV}, we shall interpret it as coming from
the high-velocity zone. Ly$\beta$ absorption has also been observed with components spanning approximately the same velocity range \citep{mirabal}. Note that this implies a very low column of {\small H\,I} in the host in front of the immediate star forming region of the GRB progenitor, unlike other GRB hosts observed to-date. We discuss these outflowing systems and their possible 
origins in turn below.

\subsection{Intervening systems, overdense regions and partial covering}
There is a possibility that the `outflow' may just be several systems in the line of sight to the burst. This, however, seems highly unlikely to occur so close to the burst redshift (see Schaefer \etal\ 2003; Fiore \etal\ 2004) and in addition to the $z=1.38$ and 1.60 systems. Specifically, a number of (proto-)galaxies within 3000~km~s$^{-1}$ of
each other would constitute a nonlinearly collapsed overdensity that
is very hard to get as early as $z\simeq2.3$ in currently favoured
cosmologies either in the Hubble flow
or within a galaxy cluster including the GRB host.

Let us now consider a clumpy stellar wind environment for GRB\,021004. Clumps of material may comprise the stellar wind -- Cherepashchuk (1990) suggested that up to 80 per cent of a Wolf-Rayet stars' mass can be lost in clumps in the wind. In principle the lines we see could
either be due to a wind with optical depth less than 1 absorbing a fraction
of the light more or less uniformly throughout the wind material, or by
a partial covering of the emitting surface by optically thick clouds.
In the latter case, the spectrum should be the sum of a featureless 
continuum from regions uncovered by cloud, and a spectrum with saturated,
flat-bottomed lines from the regions covered by clouds that are optically
thick in the lines but thin in the continuum. This implies that the summed
spectrum should have flat-bottomed lines whose flux level in the centre
is non-zero, when viewed with high enough resolution spectroscopy. Whilst
our data do not allow us to address this distinction, the UVES high resolution afterglow spectra
of GRB\,021004 (Fiore \etal\ 2004; Castro-Tirado \etal, in preparation) show flat-bottomed, saturated lines with no flux
in the centre, effectively ruling out a partial covering model.

We now consider the exotic possibility that the high-velocity absorption
features are from a number of supernova shells produced by an earlier
stellar death in the same cluster that produced the progenitor of the GRB.
It is plausible that an unusually rich cluster (e.g. the central
cluster of the S Dor complex in the Large Magellanic Cloud, LMC) would have produced a number of
stellar explosions in the past Myr, of which the supernova remnants are
now expanded beyond the cluster core size, so that our line of sight to
the next explosion will only pass through the approaching sides of the
shells and thus give rise to blueshifted features. The hydrogen richness is
explained in this case, since the ejecta could contain hydrogen, or sweep
it up. The problem is that the velocities are low enough that supernova
shells must have significantly decelerated. In so doing, they will shock and
be heated to X-ray temperatures, which makes species like {\small H\,I}, {\small Si\,IV} and {\small C\,IV}
rather unlikely to be present in those shells. The shell geometry has also been seen around Wolf-Rayet stars as shell nebulae.
Mirabal \etal\ (2003) demonstrated that optical absorption lines like those seen in GRB\,021004 can be produced by a fragmented shell nebula, under very particular conditions. The shell nebula would need to have been located at a distance from the GRB progenitor such that it received sufficient radiative acceleration from the burst to achieve the high velocities, but was not fully ionized by the GRB flux.
All things considered, we do not favour these options.

\subsection{A single star as progenitor}
The most natural explanation for the blueshifted absorption features is that they arise in a stellar wind. A 1/$r^2$ density profile is inferred from the light curve at very early times \citep{LiChev} suggesting a stellar wind circumburst medium.
A GRB progenitor must be a massive, evolved star. The highest velocity system observed is an outflow of $\sim$ 2800~km~s$^{-1}$, which could be the terminal velocity of a stellar wind and in this case implies the star is a Wolf-Rayet (WR) star (Mirabal \etal\ 2003; Schaefer \etal\ 2003). Such stars evolve to type Ib supernovae and are characterised by large mass loss rates in high velocity, optically thick stellar winds. The 3 outflow velocities observed in this afterglow could correspond to the fast WR wind (2800~km~s$^{-1}$), a slower wind from an earlier phase in the stellar lifetime (2400~km~s$^{-1}$) and the slow bubble moving into the ISM or mixing of winds from different phases (570~km~s$^{-1}$) as shown in simulations of massive star evolution by Van Marle, Langer \& Garc\'{i}a-Segura (2005). Massive star winds are predicted to show a higher optical depth in the {\small C\,IV} line than in the {\small Si\,IV} line \citep{LL} which seems to be true for these data. Line-locking is a characteristic of WR wind spectra, since the winds are radiatively driven (e.g. Puls \etal\ 1996), and this appears to be present in the WHT spectra \citep{moller}, seen as an overlap of the {\small C\,IV} doublet lines at the two highest redshifts. If, however, the only signature of the radiatively driven WR wind is the fastest absorption component, then line-locking would not be expected, and would be a coincidence in this case.   

There are two basic challenges to interpreting these lines as originating
in the stellar wind. The first is that the species causing the lines
must survive the initial blast of UV to gamma rays without becoming
fully ionized. The distance out to which the material
might be fully ionized can be estimated in two different ways, depending on the recombination
time-scale. If the recombination time is short compared to the 100~s duration of the ionizing flash \citep{discovery}, we can use ionization equilibrium as characterised by the ionization parameter $\xi = L/nr^2$.

If on the other hand
the recombination time is long, then a given
atom with cross-section $\sigma$ for ionization will simply absorb at least one
ionizing photon:
\begin{equation}
\sigma_{\rm i} \frac{E_{\rm tot}}{4 \pi r^2 <E_{\rm i}>} \ge 1
\end{equation}
This is the appropriate limit to use since a rough calculation of the recombination time for hydrogen gives a time-scale of the order of 10$^{5}$ years. This gives minimal distances for the survival of {\small Si\,IV}, {\small C\,IV} and {\small H\,I} of 5$\times$10$^{18}$, 2$\times$10$^{19}$ and 9$\times$10$^{20}$ cm respectively (using photoionization cross-sections and threshold energies from Verner \etal\ 1996 and $E_{\rm tot} = E_{\rm iso}$).
If we assume that these 3 species are co-located (because they have the same velocity structure) then the minimum distance out to which we could find them if they were ionized by the burst flux is far beyond the few pc radius of a typical stellar wind bubble. This excludes a stellar wind origin for these absorption lines if they are intercepted by the GRB flux.

The second challenge is the presence of hydrogen in the spectra with 
approximately the same velocity structure as {\small Si\,IV} and {\small C\,IV}, strongly
suggesting that significant amounts of hydrogen are present in the stellar
wind, with a column of neutral hydrogen $\ge$10$^{14}$ cm$^{-2}$. This is not usual
for a WR wind. A WR star will once have been hydrogen-rich,
and so the outer wind can have hydrogen, but its velocity structure will
not match that of the later, hydrogen-deficient wind.

WR stars in our own galaxy are hydrogen deficient and in most cases H is not detected at all. However in the SMC a large fraction, if not all single WN-type WR stars (nitrogen-rich WR stars) show hydrogen in their spectra \citep{WRpop1}. This was unexpected given the lack of H in Galactic WN-types, leading to the suggestion that the WN population in the SMC is fundamentally different from that of our own galaxy. So in the SMC, where the metallicity is low causing less severe mass loss and fewer WC-type (carbon-rich) WR stars, massive stars can remain H-rich in the WR phase. GRB host galaxies are thought to be metal-poor like the SMC, but the strength of {\small C\,IV} and {\small Si\,IV} and non-detection of nitrogen (particularly {\small N\,IV}, {\small N\,V}) in the afterglow spectra of GRB\,021004 is inconsistent with a WN-type WR star. In fact carbon produces the strongest absorption features, suggesting we are observing the wind of a WC-type WR star (e.g. Mirabal \etal\ 2003; Schaefer \etal\ 2003), thought to be the last evolutionary stage of the WR phase. On the other hand we do not detect {\small C\,III $\lambda$1247} which is strongly detected in samples of WC-types with e.g. FUSE \citep{fuse} and IUE \citep{iue}.

There is no significant variability in the absorption line EWs between our very
early spectrum at 0.49 days after burst, and those beginning 1.55 and 6.62 days
after burst (besides the {\small Al\,II} line discussed in Section 3.2). The three epochs appear to be in different `phases' of the peculiar afterglow light curve (Fig. \ref{photometry}).
The jumps in the light curve
at 0.8--1 and 2.5--3 days cannot then be shells through which the afterglow light passes, because the same absorption is seen at all epochs. Unless
the shells are located very far from the star so that the afterglow never overtakes them, which then makes it difficult to produce the fast velocities we observe.
21-cm measurements have shown {\small H\,I} shells around WR stars, e.g. WR 102 \citep{WR102}, but these have expansion velocities of $\sim$ 50~km~s$^{-1}$.
The Galactic WR star WR 3 may also have an unusually high H content if it is a single star rather than in binary with an O star, but again only rather low velocities of up to 400~km~s$^{-1}$ are observed \citep{WR3}, never reaching the 2400--2900 km s$^{-1}$ velocities we observe. 

\subsection{A binary system progenitor}
So, whilst a WR star is the obvious choice to explain the blue shifted, line-locked absorption complexes, this picture is difficult to reconcile with the large amount of neutral hydrogen observed, particularly at very high velocities.
This conundrum leaves us with few options. Either distant carbon-rich WR stars
are very different to those in our own galaxy and are able to retain a large fraction of their hydrogen, or the hydrogen originates elsewhere -- perhaps in a companion star wind.
There is the possibility that we are seeing a WR star in a close
binary with a hydrogen-rich main-sequence star. Even for an O star
companion, the momentum loss rate $\dot{M} v_{\infty}$ of the WR star
far exceeds that of the companion, and so the wind velocity structure at
distances much greater than the binary separation will be dominated by the WR star. However, a few to 10 per cent
of the mass will come from the companion and so is hydrogen rich.
When mixed into the WR wind it can cause the observed phenomenon
of H Ly$\alpha$ emission accompanying otherwise typical WR lines.
The \emph{a priori} probability that a WR star is in a binary with an O star is still a debatable number. Foellmi \etal\ (2003a) give theoretical binary frequency lower limits for WR+O systems in the Magellanic Clouds of $0.41\pm0.13$ (LMC) and $0.98\pm0.32$ (SMC), assuming a mass loss - metallicity relation of $\dot{M} \propto Z^{0.5}$ and excluding rotation effects. They find, however, significantly lower observed fractions of 30 and 40 per cent respectively (Foellmi \etal\ 2003a,b). Various measured values point towards the conclusion that the binary frequency of WR stars is identical to that of their progenitors and independent of the metallicity.

There is also the possibility that the WR star is in a binary with a
less massive star that is providing the observed hydrogen. 
The probability of a WR star having a lower-mass
companion (perhaps 1--3 M$_\odot$) is even less well known, since
such low-mass, low-luminosity companions are hard to detect.
The only evidence for the existence of such binaries is the
population of soft X-ray transients in our Galaxy, which consist of
a black hole orbited by a low-mass star in a very tight orbit.
Their ancestors must have been WR stars with a low-mass main-sequence
binary companion. It has been suggested that these systems produce a GRB
when the black hole forms \citep{msbinary}.

One necessary consequence of this H-admixing model is that probably only
within some angle from the orbital plane of the binary will H be mixed
into the outflowing wind. In the polar direction pure WR composition will
dominate. This will constrain our viewing angle to the system to being
not too far out of the orbital plane. For an O star, still having a
relatively strong wind, this is not severe, but for a low-mass star with
a very weak wind, the system would have to be viewed at or close to edge-on. This is unlikely if, as predicted by the collapsar model, the jets emerge from the polar axis of the progenitor star.

\begin{figure}
\begin{center}
\includegraphics[width=12cm]{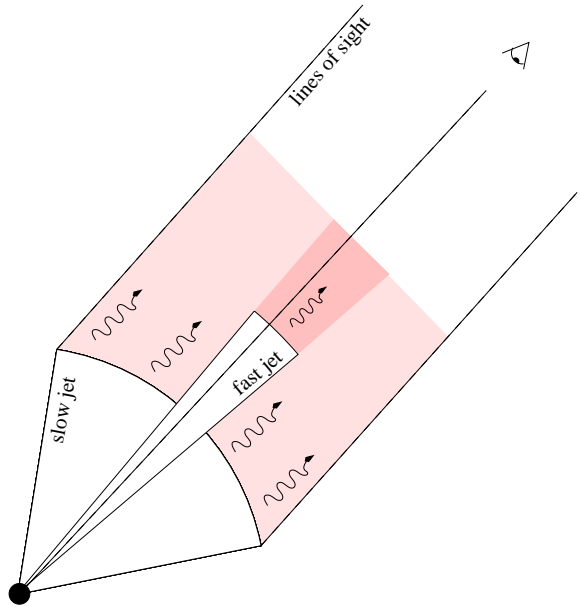}
\caption{A sketch of the structured jet geometry within the stellar wind, where our lines-of-sight (dashed lines) intersect both fully ionized material infront of the narrow, fast jet (dark grey) and material ionized to a much lesser degree by the the slower jet (light grey).}
\label{sketch}
\end{center}
\end{figure}
\subsection{A structured jet model}
The ionization problem remains however, and it seems that in order to observe fast-moving species such as {\small Si\,IV} and {\small C\,IV} and even {\small H\,I}, one must be observing a region undisturbed by the gamma ray jet. In order to achieve this we suggest a structured jet geometry, with one narrow jet ($\lesssim$ 3$^{\circ}$) and one slower, much wider jet of material which are both visible to us (Fig. \ref{sketch}). In fact, the afterglow has a surface area, and so there are several lines-of-sight along which the light will reach us. Gamma rays from the inner jet will ionize material in its path. But if the inner jet opening angle is small then there will also be a line-of-sight along which we see material unaffected by high energy radiation. We propose that there is a second wider cone of slower jet material surrounding the narrow gamma ray jet. This wide jet is also seen through the absorber, probably a fast stellar wind within 2 older more slowly moving wind phases, which produces the absorption lines we observe. 

The power law decay of afterglow light curves can be very smooth, as in the cases of GRB\,020813 (e.g. Laursen \& Stanek 2003) and GRB\,990510 (e.g. Stanek \etal\ 1999). But some show considerable deviations from this with many irregularities, as seen in GRB\,011211 (e.g. Holland \etal\ 2002) and GRB\,030329 (e.g. Lipkin \etal\ 2004) and this is the case for this burst (Fig \ref{photometry}).
The structured jet scenario we describe is consistent with the observed light curve of the afterglow of GRB\,021004 (Van der Horst, private communication).
We also note that the general decay of the light curve may be adequately fit with other models including several energy injections \citep{icelandic}, patchy shell models \citep{patchyshell} and density enhancements in the ISM \citep{Lazz} or in a wind \citep{heylperna}.

From the light curve alone we cannot pinpoint the cause of the irregularities in the temporal decay. Considering the light curve in combination with the spectra provides further constraints and leads us to prefer the structured jet idea, which can explain both types of observation and does not require the absorbers to be at a large distance from the progenitor to produce the absorption lines seen in our spectra.

\section{Conclusions}
We have observed the optical afterglow of GRB\,021004 spectroscopically at
three epochs ranging from 0.49 to 6.62 days after the GRB, when the emitting
surface is of order $10^{16}$~cm away from the explosion centre. The most
unique feature in the spectra is the fact that lines of {\small H\,I}, {\small Si\,IV} and {\small C\,IV} are observed at blueshifts of up to 2900~km~s$^{-1}$.
While no model explored by ourselves or earlier authors explains this phenomenon
in a very simple manner, we find that by far the most plausible 
explanation is that these lines result in the fossil stellar wind of a highly
evolved Wolf-Rayet star. It is unusual that hydrogen would be present in such a
wind. This could be explained if WR stars do remain hydrogen-rich up to explosion at high redshift and/or low metallicity,
or if a nearby companion star had enriched the WR
wind with hydrogen. In any case, our data seem to further enhance the case
for the origin of GRBs in the core collapse of very massive stars.
Ionization from the burst itself prevents the existence of {\small H\,I}, {\small Si\,IV} and {\small C\,IV} close to the afterglow surface where the fast stellar wind should dominate. A structured jet geometry is a promising explanation: the gamma rays emerge in a small opening angle leaving a wider cone of slower, less energetic material also along our line-of-sight. This wider jet does not fully ionize the material in its path, and can give rise to the absorption lines we observe. This model is able to explain both the spectral features and reproduce the light curve shape.   

\section{Acknowledgments}
We thank R. McLure for interrupting his observing run to take these WHT data. We acknowledge useful discussions with C. Dijkstra, M. Franx, A. J. van der Horst, A. de Koter, A.-J. van Marle and K. Wiersema.
The WHT is operated on the island of La Palma by the Isaac Newton Group in the Spanish Observatorio del Roque de los Muchachos of the Instituto de Astrofisica de Canarias. The VLT data were obtained under ESO proposal number 70.D-0523(A). The authors acknowledge benefits from collaboration within the Research Training Network
`Gamma-Ray Bursts: An Enigma and a Tool', funded by the EU under contract
HPRN-CT-2002-00294.

\bsp
\label{lastpage}

\end{document}